\documentclass[11pt,a4paper,twoside,groupcitations]{article}
\usepackage[T1]{fontenc}
\usepackage[ansinew]{inputenc}
\usepackage[english]{babel}
\usepackage{amsfonts}
\usepackage{amsmath}
\usepackage{bm}
\usepackage{array}
\usepackage{amsthm}
\usepackage{amssymb}
\usepackage{graphicx}
\usepackage{subfigure}
\usepackage{braket}
\usepackage{eucal}
\usepackage{verbatim}
\usepackage[table]{xcolor}
\usepackage{caption}
\usepackage{cite}
\usepackage{textcomp}
\usepackage{multicol}
\usepackage{tikz}
\usetikzlibrary{positioning,arrows}
\usetikzlibrary{decorations.pathmorphing}
\usetikzlibrary{decorations.markings}
\usetikzlibrary{calc,decorations.markings}
\usetikzlibrary{arrows,shapes}
\usetikzlibrary{matrix,arrows}
\usepackage{pgfplots}
\usepackage{xparse}
\definecolor{jade}{HTML}{00A86B}
\newcommand{\be}{\begin{eqnarray}}
\newcommand{\ee}{\end{eqnarray}}

\newcommand{\expec}[1]{\mbox{$\langle\, #1\,\rangle$}}

\renewcommand{\a}{\hat a}
\newcommand{\ac}{\hat a^{\dagger}}

\renewcommand{\d}{\mbox{${\rm d}$}} 
\newcommand{\lp}{\ell_{\rm p}}
\newcommand{\mpl}{m_{\rm p}}
\newcommand{\gn}{G_{\rm N}}
\newcommand{\rh}{r_{\rm H}}
\newcommand{\Rh}{R_{\rm H}}

\newcommand{\Ng}{{N_{\rm G}}}

\newcommand{\Rinf}{{R_{\infty}}}
\newcommand{\dd }{{\mathrm d}}
\title{\bf Geometry and thermodynamics of coherent quantum black holes}
\author{Roberto~Casadio$^{ab}$\thanks{E-mail: casadio@bo.infn.it}
\\
\\
$^a${\em Dipartimento di Fisica e Astronomia, Universit\`a di Bologna}
\\
{\em via Irnerio~46, 40126 Bologna, Italy}
\\
\\
$^b${\em I.N.F.N., Sezione di Bologna, I.S.~FLAG}
\\
{\em viale B.~Pichat~6/2, 40127 Bologna, Italy}
}
\begin{document}
\maketitle
\begin{abstract}
We present a quantum description of black holes given by coherent states of gravitons
sourced by a matter core.
The expected behaviour in the weak-field region outside the horizon is recovered,
with arbitrarily good approximation, but the classical central singularity is not resolved
because the coherent states may not contain modes of arbitrarily short wavelength and
the matter core must therefore have finite size.
Ensuing quantum corrections both in the interior and exterior are also estimated
by assuming the mean-field approximation holds everywhere. 
These deviations from the classical black hole geometry can be viewed as 
quantum hair and lead to a quantum corrected horizon radius and thermodynamics.
\end{abstract}
\section{Introduction and motivation}
\setcounter{equation}{0}
\label{S:intro}
The gravitational collapse of compact objects generates geodesically
incomplete spacetimes in general relativity if a trapping surface appears~\cite{HE}
and eternal point-like sources are not mathematically compatible with Einstein's field
equations~\cite{geroch}.
We expect the quantum theory will fix this inconsistent classical picture of the gravitational interaction, 
in the same way that quantum mechanics explains the stability of atoms by not admitting quantum states
corresponding to the classical ultraviolet catastrophe.
Although the fundamental dynamics remains unaffected, the condition for the existence of a quantum
state can greatly modify the effective description of any systems.  
At the same time, the quantum state of a macroscopic black hole must be such that the effective
description reproduces the phenomenology of spacetime we detect
experimentally~\cite{Goddi:2019ams,LIGOScientific:2018mvr}.
\par
There are many quantum models of black holes in the literature (for a very partial list,
see Refs.~\cite{DvaliGomez,qbh1,qbh2}). 
In particular, the corpuscular picture~\cite{DvaliGomez} belongs to the class of approaches for which geometry
should only emerge at suitable (macroscopic) scales from the underlying (microscopic) quantum field theory of
gravitons~\cite{feynman,deser}.
The key idea is that the constituents of black holes are soft gravitons marginally bound in their
own potential and forming a condensate~\cite{DvaliGomez} with characteristic Compton-de~Broglie
wavelength $\lambda_{\rm G}\sim \Rh$, where the gravitational (or Schwarzschild) radius of the black hole of
Arnowitt-Deser-Misner (ADM) mass~\cite{ADM} $M$ is given by~\footnote{Units with $c=1$ are used throughout
and the Newton constant $\gn=\lp/\mpl$, where $\lp$ is the Planck length and $\mpl$ the Planck mass, so that $\hbar=\lp\,\mpl$.}
\be
\Rh=2\,\gn\,M
\ .
\ee
The energy scale of the gravitons is correspondingly given by $\epsilon_{\rm G}\sim \hbar/\lambda_{\rm G}$ and,
if one assumes that the total mass of the black hole $M\simeq N_{\rm G}\,\epsilon_{\rm G}$,
there immediately follows the scaling relation for the total number of gravitons
\be
N_{\rm G}
\sim 
\frac{M^2}{\mpl^2}
\sim
\frac{\Rh^2}{\lp^2}
\ ,
\label{Ng}
\ee
which reproduces Bekenstein's conjecture for the horizon area quantisation~\cite{bekenstein}.
\par
The nonlinearity of the gravitational interaction plays a crucial role in this picture.
This can be seen by considering that the (negative) gravitational energy of a source of mass $M$
localised inside a sphere of radius $R_{\rm s}$ is given by $U_{\rm N}\sim M\, V_{\rm N}(R_{\rm s})$,
where 
\be
{V}_{\rm N}
=
-\frac{\gn\,M}{r}
\label{Vn}
\ee
here is the Newtonian potential.
This potential can be obtained as the expectation value of a scalar field on
a coherent state, whose normalisation then yields the graviton number~\eqref{Ng}
for any values of $R_{\rm s}\gtrsim\Rh$~\cite{Casadio:2016zpl,Casadio:2017cdv,Mueck:2013mha,Bose:2021ytn}.
In addition to that, assuming most gravitons have the same wavelength $\lambda_{\rm G}$,
the binding energy of each graviton is given by
\be
\epsilon_{\rm G}
\sim
\frac{U_{\rm N}}{N_{\rm G}}
\sim
-\frac{\lp\,\mpl}{R_{\rm s}} 
\ ,
\ee
which yields the typical Compton-de~Broglie length $\lambda_{\rm G}\sim R_{\rm s}$.
The graviton self-interaction energy hence reproduces the (positive) post-Newtonian energy,
\be
U_{\rm GG}
\sim
N_{\rm G} \, \epsilon_{\rm G} \, V_{\rm N}(R_{\rm s})
\sim
\frac{\gn^2\,M^3}{R_{\rm s}^2}
\ ,
\label{Ugg}
\ee
and the fact that gravitons in a black hole are marginally bound~\cite{DvaliGomez}, that is
$U_{\rm N}+U_{\rm GG}\simeq 0$, finally yields the scaling
$\lambda_{\rm G}\sim R_{\rm s}\simeq \Rh$~\cite{Casadio:2016zpl,Casadio:2017cdv}.
\par
A key feature of the above scenario is that, like for all bound states in quantum physics,
it does not contain modes of arbitrarily small wavelengths and the classical central 
singularity is therefore not realised.~\footnote{This assumption
can be viewed as a manifestation of the classicalization of gravity~\cite{classicalization}.
Independent arguments in support of this condition, based on the quantum nature of the source,
were also given in Refs.~\cite{Casadio:2020ueb,Casadio:2021cbv}.
For other similar considerations, see Refs.~\cite{hofmann}.}
However, viewing a black hole as a quantum state made of only gravitons with one
typical wavelength $\lambda_{\rm G}$ cannot reproduce the gravitational field in the
accessible outer spacetime, even in the simple Newtonian approximation.
Of course, general relativity is a metric theory and a complete description of quantum gravity
remains beyond our scope, but this (conceptually and phenomenologically)
important issue can be addressed in a simplified form for static and spherically symmetric 
systems~\cite{Casadio:2016zpl,Casadio:2021onj,Giusti:2021shf,Casadio:2022ndh}.
In particular, we limit the quantum description to the potential function~\eqref{Vn} appearing in the
Schwarzschild metric
\be
\d s^2
=
-\left(1+2\,V_{\rm N}\right)\d t^2
+
\frac{\d r^2}{1+2\,V_{\rm N}}
+
r^2\,\d\Omega^2
\ .
\label{gSch}
\ee
In fact, the function $V_{\rm N}$ plays the role of a potential in the geodesic equation of radial motion,
\be
\frac{\d^2 r}{\d\tau^2}
\equiv
\ddot r
=
-\frac{\gn\,M}{r^2}
=
-V_{\rm N}'
\ ,
\label{geodesic}
\ee
where the geodesic $r=r(\tau)$ is parameterised by the proper time $\tau$,~\footnote{One should
also notice that the general relativistic function $V_{\rm N}$ depends
on the areal radius $r$, whereas the post-Newtonian viewpoint in Eqs.~\eqref{Vn}-\eqref{Ugg}
makes use of the harmonic radius~\cite{weinberg,Casadio:2021cbv,Casadio:2021gdf}.}
and this equation is the starting point to derive a discrete spectrum for a collapsing dust core~\cite{Casadio:2021cbv}.
\par
The first important step in a field theoretic description of gravity is to identify the quantum
vacuum $\ket{0}$, which we assume is the realisation of a universe where no modes (of matter or gravity)
are excited.
We could then associate the Minkowski metric $\eta_{\mu\nu}$ to such an
absolute vacuum, since this is the metric used to describe linearised
gravity and to define both matter and gravitational excitations in this regime.
The linearised theory should provide a reliable description for small matter sources
(say with total energy $M\ll\mpl$), for which it allows one to recover the Newtonian
potential from simple tree-level graviton exchanges in the non-relativistic limit.
In this context, the potential $V_{\rm N}$ is not a fundamental scalar, but it emerges from
the (non-propagating) longitudinal polarisation of virtual gravitons~\cite{feynman}.
For large sources (that is, with $M\gg\mpl$), one can in principle reconstruct the complete
classical dynamics~\cite{deser},~\footnote{See Ref.~\cite{Padmanabhan:2004xk} for a critical discussion
of this conjecture.}
but to obtain the proper quantum state from the excitations
of the linearised theory appears rather hopeless in this highly non-linear regime.
In order to circumvent this issue and, at the same time, to compare with experimental data,
one usually assumes that there exists a classical background geometry to replace $\eta_{\mu\nu}$
with a suitable solution $g_{\mu\nu}$ of the corresponding classical Einstein's equations.
However, in our perspective, the Einstein equations should emerge as effectively describing the
macroscopic dynamics and a mean-field metric should correspond to a suitable quantum state.
Since we are interested in static and spherically symmetric configurations representing a black hole,
we just require that the relevant quantum state of gravity effectively reproduces (as closely as possible)
the expected Schwarzschild geometry~\eqref{gSch} like it was done in Ref.~\cite{Casadio:2021onj}.
This approach is thus akin to quantising the longitudinal mode of gravity~\cite{feynman,Christodoulou:2022knr} 
and can be obtained by employing coherent states in a suitable Fock space built upon the Minkowski vacuum.
The use of coherent states is generically motivated by their property of minimising the quantum uncertainty,
and is further supported by studies of electrodynamics~\cite{Mueck:2013mha,Muck:2013orm},
linearised gravity~\cite{Bose:2021ytn} and the de~Sitter spacetime~\cite{Berezhiani:2021zst}.
\par
In the original corpuscular picture, as briefly reviewed above, baryonic matter sourcing the gravitational
field and triggering the gravitational collapse is argued to become essentially irrelevant
after the black hole forms~\cite{DvaliGomez}.
In Refs.~\cite{Casadio:2021cbv,Casadio:2021onj}, we instead found that a finite size $R_{\rm s}$
of the matter source is both a consequence of the quantum nature of collapsing matter and a necessary condition
for the existence of a proper quantum state which (approximately) reproduces the outer Schwarzschild geometry.~\footnote{In a fully
quantum picture, one expects that $R_{\rm s}$ be at best the expectation value of some emergent operator in the relevant state
of matter fields.  
For more considerations on the scale $R_{\rm s}$ for quantum black holes, see the concluding remarks
and Refs.~\cite{Casadio:2021cbv,Almeida:2021sci}.}
This supports the point of view of Refs.~\cite{Casadio:2016zpl,Casadio:2017cdv,Casadio:2020ueb,Casadio:2019tfz},
according to which matter inside the black hole still plays a very significant role
in defining the structure of the interior of astrophysical black holes (and possibly microscopic ones)
and leads to the existence of quantum hair~\cite{Calmet:2021stu}.
In fact, the quantum version of $V_{\rm N}$ can be employed in order to reconstruct a quantum corrected 
complete metric~\cite{Casadio:2021onj}.
In this work, we analyse in details the quantum corrected black hole geometry and thermodynamics,
thus complementing the study of the entropy of the matter core of Ref.~\cite{Casadio:2022pla}.  
\par
In the next section, we will briefly review how coherent states of a massless scalar field on a
reference flat spacetime can be used to reproduce a classical static field configuration, in general;
a consistent quantum state for the metric function~\eqref{Vn} is then introduced 
in Section~\ref{S:QBH}, where the corresponding geometry and thermodynamics will be analysed
in details;
final remarks, outlook and connections with other works will be given in Section~\ref{S:conc}.
\section{Quantum coherent states for classical static configurations}
\label{coherent}
\setcounter{equation}{0}
We will first review how to describe a generic static $V=V(r)$ as the mean field of the coherent state of
a free massless scalar field (for more details, see, {\em e.g.}~Ref.~\cite{Mueck:2013mha,Casadio:2017cdv}).
It is important to remark again that this description is not fundamental and the use of a scalar field to represent
the (non-perturbative) behaviour of the true degrees of freedom of Einstein gravity in the (rather unphysical)
static limit is a simplification supported by more refined analyses~\cite{Bose:2021ytn,Berezhiani:2021zst}.
\par
We first rescale the dimensionless $V$ so as to obtain a canonically normalised real scalar field
$\Phi =\sqrt{{\mpl}/{\lp}}\, V$, and then quantise $\Phi$ as a massless field satisfying
the free wave equation
\be
\left[
-\frac{\partial^2}{\partial t^2}
+
\frac{1}{r^2}\,\frac{\partial}{\partial r}
\left(r^2\,\frac{\partial}{\partial r}\right)
\right]
\Phi(t,r)
\equiv
\left(-\partial_t^2+\triangle\right)
\Phi
=
0
\ .
\label{KG}
\ee
Solutions to Eq.~\eqref{KG} can be conveniently written as
\be
u_{k}(t,r) = e^{-i\,k\,t}\,j_0(k\,r)
\ ,
\label{u_k}
\ee
where $k>0$ and $j_0={\sin(k\,r)}/{k\,r}$
are spherical Bessel functions satisfying the orthogonality relation 
\be
4\,\pi
\int_0^\infty
r^2\,\dd r\,
j_0(k\,r)\,j_0(p\,r)
=
\frac{2\,\pi^2}{k^2}\,
\delta(k-p)
\ .
\ee
The quantum field operator and its conjugate momentum read
\be
\label{Phi}
\hat{\Phi}(t,r)
\!\!&=&\!\! 
\int_0^\infty
\frac{k^2\,\dd k}{2\,\pi^{2}}\,
\sqrt{\frac{\hbar}{2\,k}}
\left[
\a_{k}\,
u_k(t,r)
+
\ac_{k}\, 
u^*_k(t,r)
\right]
\\
\hat{\Pi}(t,r) 
\!\!&=&\!\!
i\int_0^\infty
\frac{k^2\,\dd k}{2\,\pi^{2}}\,
\sqrt{\frac{\hbar\,k}{2}}
\left[
\a_{k}\,
u_k(t,r)
-
\ac_{k}\, 
u^*_k(t,r)
\right]
\ ,
\ee
which satisfy the equal time commutation relations,
\be
\left[\hat{\Phi}(t,r)
,\hat{\Pi}(t,s)\right] 
=
\frac{i\,\hbar}{4\,\pi\,r^2}\,
\delta(r-s)
\ ,
\ee
provided the creation and annihilation operators obey the commutation rules
\be
\left[\a_{k},\ac_{p}\right]
=
\frac{2\,\pi^2}{k^2}\,
\delta(k-p)
\ .
\ee
The Fock space of quantum states is then built from the vacuum defined by $\a_{k}\ket{0}=0$
for all $k>0$.
The choice of the flat Minkowski metric in Eq.~\eqref{KG} follows from the expectation that
this vacuum $\ket{0}$ is meant to describe a completely empty spacetime devoid of any matter source
and without excitations of the gravitational field, as we argued in the introductory Section~\ref{S:intro}.
\par
Classical configurations of the scalar field that can be realised in the quantum theory
must correspond to suitable states in this Fock space,
and a natural choice is given by coherent states $\ket{g}$ such that
\be
\a_{k} \ket{g} = g_{k}\,e^{i\,\gamma_{k}(t)} \ket{g}
\ .
\ee
In particular, we are interested in those $\ket{g}$ for which the expectation value of the quantum
field $\hat{\Phi}$ reproduces the classical potential, namely
\be
\sqrt{\frac{\lp}{\mpl}}
\bra{g}\hat{\Phi}(t,r)\ket{g}
=
V(r)
\ .
\label{expecphi}
\ee
From the expansion~\eqref{Phi}, we obtain
\be
\bra{g}\hat{\Phi}(t,r)\ket{g}
=
\int_0^\infty
\frac{k^2\,\dd k}{2\,\pi^2}\,
\sqrt{\frac{2\,\lp\,\mpl}{k}}\,
g_k\,
\cos[\gamma_k(t)-k\,t]\,
j_0(k\,r)
\ .
\ee
If we now write 
\be
V
=
\int_0^\infty
\frac{k^2\,\dd k}{2\,\pi^2}\,
\tilde V(k)\,
j_0(k\,r)
\ ,
\label{Vk}
\ee
we immediately obtain~\footnote{This formal choice of phases eliminates the time dependence from the
normal modes~\eqref{u_k} and allows for reproducing static configurations.
From a physical point of view, one can consider such a limiting approximation holds for time intervals
shorter than $\Delta t\sim k^{-1}$.} 
\be
\gamma_k=k\,t
\ee
and
\be
g_k
=
\sqrt{\frac{k}{2}}\,
\frac{\tilde V(k)}{\lp}
\ .
\label{gkVk}
\ee
The coherent state finally reads
\be
\ket{g}
=
e^{-N_{\rm G}/2}\,
\exp\left\{
\int_0^\infty
\frac{k^2\,\dd k}{2\,\pi^2}\,
g_k\,
\ac_k
\right\}
\ket{0}
\ ,
\label{gstate}
\ee
where
\be
\label{NGN}
\Ng
=
\int_0^{\infty} 
\frac{k^2\,\dd k}{2\,\pi^2}\, 
g_k ^2
\ee
is the total occupation number.
We note in particular that the value of $\Ng$ measures the ``distance'' in the Fock Space of $\ket{g}$
from the vacuum $\ket{0}$ corresponding to $\Ng=0$.
Another quantity of interest is given by 
\be
\label{EN}
\expec{k}
=
\int_0^{\infty} 
\frac{k^2\,\dd k}{2\,\pi ^2} 
\,k\,g_k ^2
\ ,
\ee
from which one obtains the ``average'' wavelength $\lambda_{\rm G}=\Ng/\expec{k}$.
\section{Quantum Schwarzschild black holes}
\label{S:QBH}
\setcounter{equation}{0}
We will now apply results from the previous section to the Schwarzschild metric~\eqref{gSch}.
This geometry contains only the function $V_{\rm N}=\sqrt{\gn}\,\Phi$, and
all of the relevant expressions introduced can be explicitly computed
from the coefficients $g_k$ representing the occupation numbers of the modes $u_k$.
\par
In particular, by inverting Eq.~\eqref{Vk}, we find
\be
\tilde{V}_{\rm N}
=
-4\,\pi\,\gn\,\frac{M}{k^2}
\ee
and the coefficients
\be
g_k
=
-\frac{4\,\pi\,M}{\sqrt{2\,k^3}\,\mpl}
\ .
\label{gkN}
\ee
For such a coherent state, we obtain
\be
N_{\rm G}
=
\frac{4\,M^2}{\mpl^2}
\int_0^\infty
\frac{\dd k}{k}
\ee
and
\be
\expec{k}
=
\frac{4\,M^2}{\mpl^2}
\int_0^\infty
\dd k
\ .
\ee
The number of quanta $N_{\rm G}$ contains a logarithmic divergence both in the infrared (IR) and
the ultraviolet (UV), whereas $\expec{k}$ only diverges (linearly) in the UV.
\par
\begin{figure}[t]
\centering
\includegraphics[width=8cm]{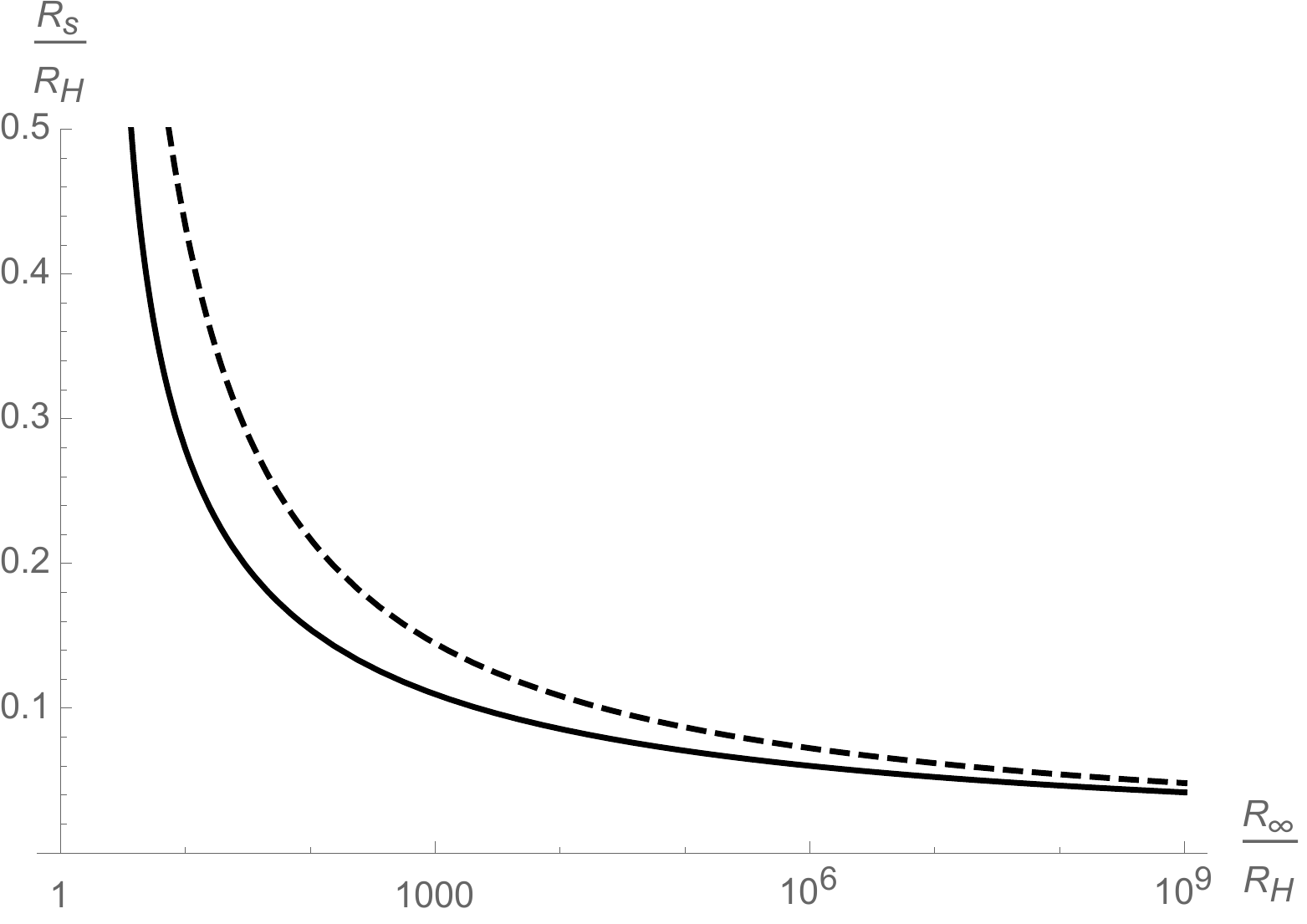}
\caption{Exact solution to Eq.~\eqref{eq:RsRh} (solid line) compared to its approximation~\eqref{app:RsRh}
(dashed line).}
\label{RsRh}
\end{figure}

The meaning of such divergences was already explored in details in previous
works~\cite{Casadio:2017cdv,Casadio:2020ueb,Casadio:2021onj}.
In particular, we recall that the UV divergences are due to the vanishing size of
the source, hence they would not be present if the density were regular.
Instead of smoothing out the source, the UV divergences can be formally
regularised by introducing a cut-off $k_{\rm UV}\sim 1/R_{\rm s}$, where $R_{\rm s}$ 
can be interpreted as the finite radius of a would-be-regular matter source.
Such a cut-off is just a mathematically simple way of describing the fact that the very existence of a proper
quantum state $\ket{g}$ requires the coefficients $g_k$ to depart from their purely classical expression~\eqref{gkN}
for $k\to 0 $ and $k\to\infty$.
Of course, for a black hole spacetime, we must have $R_{\rm s}<\Rh$.
Likewise, we introduce a IR cut-off $k_{\rm IR} = 1/\Rinf$ to account for the necessarily finite
life-time $\tau\sim \Rinf$ of a realistic source~\cite{Casadio:2017cdv}~\footnote{Any disturbance
in the source will propagate (at most) at the speed of light.
Alternatively, we could take for $R_\infty$ the size of the observable Universe as an upper bound~\cite{Giusti:2021shf}.
In any case, modes with wavelength $k^{-1}$ many orders of magnitude larger than $\Rh$ do not contribute
significantly to the determination of $V=V(r)$ in Eq.~\eqref{expecphi}
(for the details, see Ref.~\cite{Casadio:2020ueb,DvaliSoliton}) and we shall assume $k_{\rm IR}\to 0$
whenever possible.}
and rewrite
\be
N_{\rm G}
=
\frac{4\,M^2}{\mpl^2}
\int_{k_{\rm IR}}^{k_{\rm UV}}
\frac{\dd k}{k}
=
4\,\frac{M^2}{\mpl^2}\,
\ln\left(\frac{R_\infty}{R_{\rm s}}\right)
\label{cNg}
\ee
and
\be
\expec{k}
=
\frac{4\,M^2}{\mpl^2}
\int_{k_{\rm IR}}^{k_{\rm UV}}
\dd k
=
4\,\frac{M^2}{\mpl^2}
\left(
\frac{1}{R_{\rm s}}
-
\frac{1}{R_\infty}
\right)
\ .
\label{ckg}
\ee
The corpuscular scaling~\eqref{Ng} for the number $N_{\rm G}$ with the square of the energy $M$ of the system  
already appears at this stage, whereas the second crucial result
\be
\lambda_{\rm G}
=
\frac{N_{\rm G}}{\expec{k}}
\sim 
\lp\,\frac{M}{\mpl}
\ ,
\ee
is obtained from Eqs.~\eqref{cNg} and \eqref{ckg} only provided the cut-offs satisfy
\be
\ln\left(\frac{R_\infty}{R_{\rm s}}\right)
\simeq 
\frac{\Rh}{R_{\rm s}}
\ .
\label{eq:RsRh}
\ee
Assuming $R_{\rm s}\lesssim \Rh\ll R_\infty $, the above yields (see Fig.~\ref{RsRh} for a graphical
comparison with the exact solution)
\be
R_{\rm s}
\simeq
\frac{\Rh}{\ln\left({R_\infty}/{\Rh}\right)}
\ ,
\label{app:RsRh}
\ee
so that the size of the inner source and the radius of the outer region containing a gravitational field appear connected
in the quantum description.
We will further comment about possible consequences of this result in the concluding Section~\ref{S:conc}.
\subsection{Consistent black hole states}
We started from the condition in Eq.~\eqref{expecphi}, which demands that the coherent state $\ket{g}$
reproduces the classical potential everywhere.
We then found that acceptable occupation numbers $g_k$ do not exist which satisfy this requirement for 
$k\to 0$ and $k\to\infty$. 
We next observe that, for a black hole, the mean field needs only reproduce the classical $V_{\rm N}$
with sufficient accuracy to comply with experimental bounds at most in the region of outer communication
outside the horizon.
This means that the coherent state $\ket{g_{\rm BH}}$ representing a black hole must give
\be
\sqrt{\frac{\lp}{\mpl}}
\bra{g_{\rm BH}}\hat{\Phi}(t,r)\ket{g_{\rm BH}}
\simeq
V_{\rm N}(r)
\qquad
{\rm for}
\
r\gtrsim \Rh
\ ,
\label{QCo}
\ee
where we recall that $V_{\rm N}(\Rh)=-1/2$ and the approximate equality is subject to experimental precision.
In practice, this weaker condition means that $\ket{g_{\rm BH}}$ does not need to contain the modes of infinitely
short wavelength that are necessary to resolve the classical singularity at $r=0$.
\par
\begin{figure}[t]
\centering
\includegraphics[width=10cm]{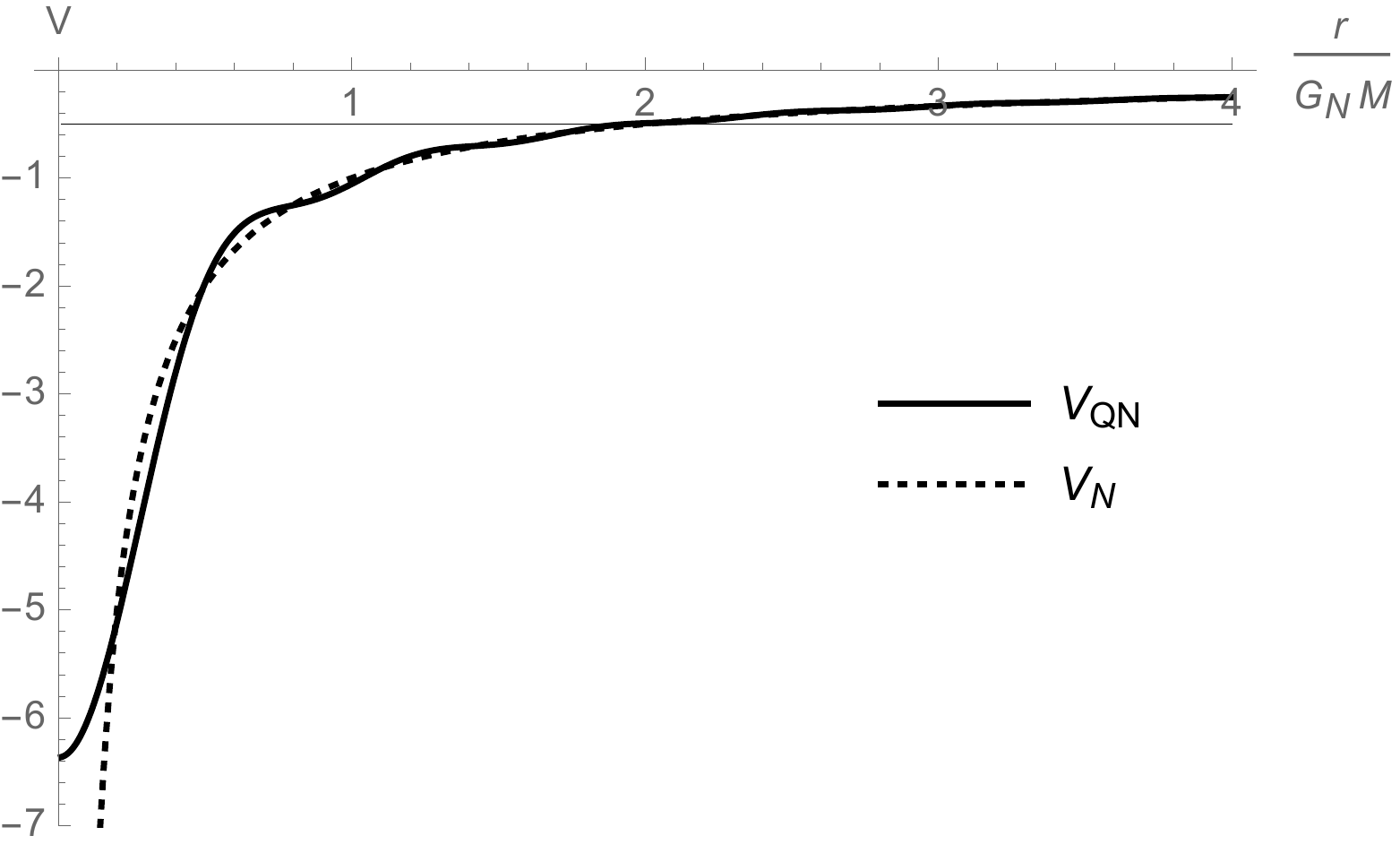}
\caption{Quantum metric function $V_{\rm QN}$ in Eq.~\eqref{Vq} (solid line) compared to $V_{\rm N}$
(dashed line) for $R_{\rm s}=\Rh/20$.
The horizontal thin line marks the location of the horizon for $V=-1/2$.}
\label{VqVn}
\end{figure}

In fact, Eq.~\eqref{QCo} can be satisfied by building the coherent state $\ket{g_{\rm BH}}$ according
to Eq.~\eqref{gstate} with modes of wavelength $k^{-1}$ larger than some fraction of the size
of the gravitational radius $\Rh$ of the source, which we can further identify with the UV cut-off $R_{\rm s}$.
By momentarily considering also the IR scale $k_{\rm IR}$, we thus have that only the modes $k$ satisfying
\be
R_\infty^{-1}
\sim
k_{\rm IR}
\lesssim 
k
\lesssim
k_{\rm UV}
\sim
R_{\rm s}^{-1}
\ee
are significantly populated in the quantum state $\ket{g_{\rm BH}}$.
This yields an effective quantum potential
\be
V_{\rm QN}
&\!\!\simeq\!\!&
\int_{k_{\rm IR}}^{k_{\rm UV}}
\frac{k^2\,\dd k}{2\,\pi^2}\,
\tilde V_{\rm N}(k)\,j_0(k\,r)
\nonumber
\\
&\!\!\simeq\!\!&
-
\frac{2\,\lp\,M}{\pi\,\mpl\,r}
\int^{r/R_{\rm s}}_{0}
\dd z\,
\frac{\sin z}{z}
\ ,
\label{Vqq}
\ee
where we defined $z=k\,r$ and let $k_{\rm IR}=1/R_\infty\to 0$ as mentioned above.
We thus find
\be
V_{\rm QN}
&\!\!\simeq\!\!&
-\frac{2\,\gn\,M}{\pi\,r}\,
{\rm Si}\left(\frac{r}{R_{\rm s}}\right)
\nonumber
\\
&\!\!\simeq\!\!&
V_{\rm N}
\left\{
1
-\left[1-
\frac{2}{\pi}\,{\rm Si}\left(\frac{r}{R_{\rm s}}\right)
\right]
\right\}
\ ,
\label{Vq}
\ee
where ${\rm Si}$ denotes the sine integral function (see Fig.~\ref{VqVn} for an example).
\par
It is important to remark that different shapes for the deviation from the classical potential $V_{\rm N}$
would be obtained if one employed a different UV cut-off in the integral~\eqref{Vqq}.
In particular, one could consider a smooth window function rather than the hard cut-off $k<R_{\rm s}^{-1}$.
Assuming such a smooth function is physically related to the distribution of matter in the black hole interior,
one could in principle detect different matter profiles from analysing (test particle motion in)
the black hole exterior, whose geometry is going to be described in details next.
However, we shall not endeavour in the survey of other possibilities in the present work,
as very little can be accomplished analytically.
\subsection{Geometry}
\label{A:geo}
\begin{figure}[t]
\centering
\includegraphics[width=10cm]{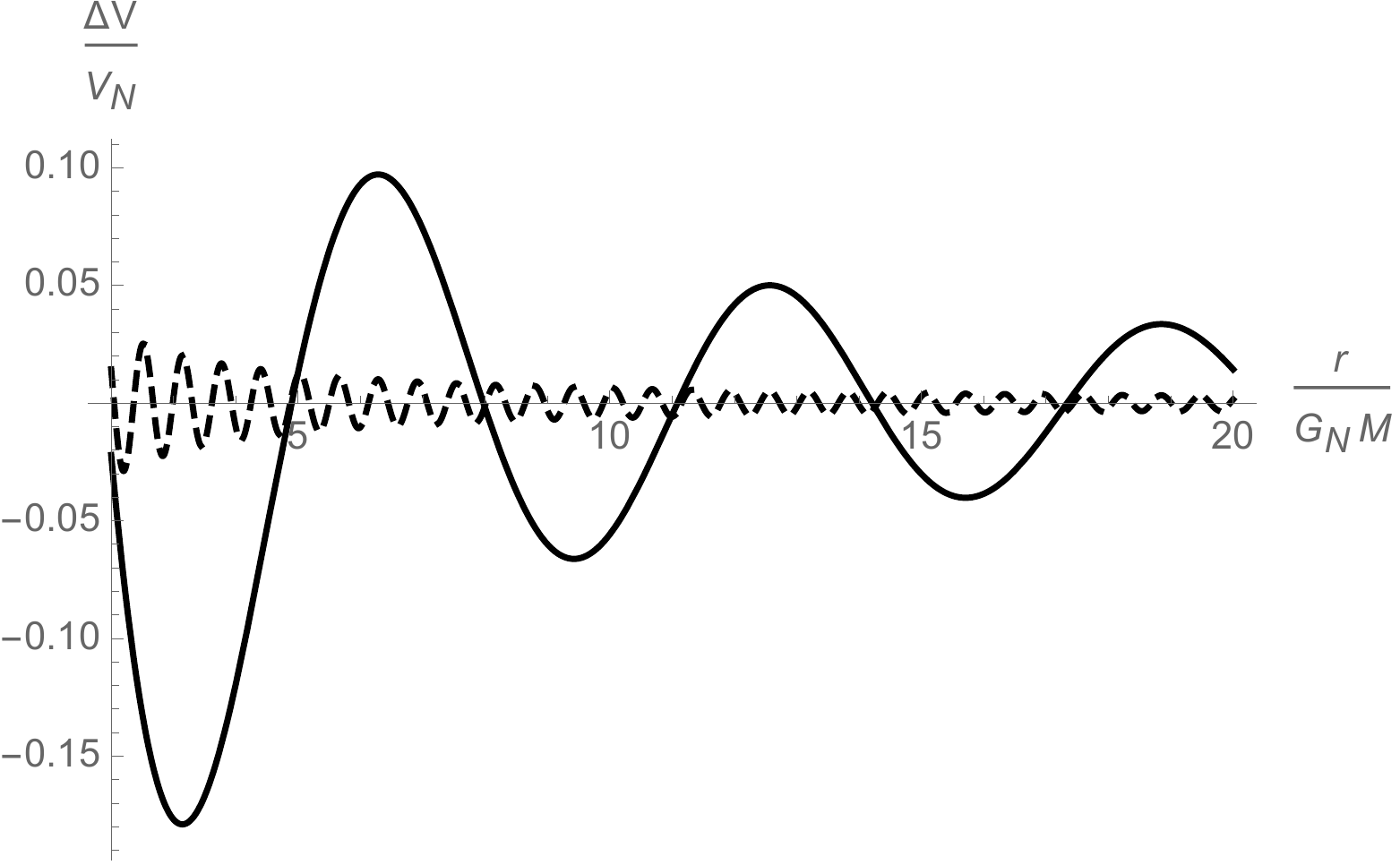}
\caption{Oscillations of the quantum potential $V_{\rm QN}$ in Eq.~\eqref{Vq} around the Schwarzschild
expression $V_{\rm N}$ for $R_{\rm s}=\gn\,M=\Rh/2$ (solid line) and
$R_{\rm s}=\Rh/20$ (dashed line) in the region outside 
the horizon $\Rh=2\,\gn\,M$.}
\label{dVq}
\end{figure}
The potential~\eqref{Vq} can be used to reconstruct the full spacetime metric straightforwardly as~\cite{Casadio:2021onj}
\be
\d s^2
=
-\left(1+2\,V_{\rm QN}\right)\d t^2
+
\frac{\d r^2}{1+2\,V_{\rm QN}}
+
r^2\,\d\Omega^2
\ ,
\label{gQ}
\ee
where the dependence of $V_{\rm QN}$ on $R_{\rm s}$ therefore results in a quantum violation
of the no-hair theorem~\cite{Calmet:2021stu}.
Near the origin, this quantum corrected metric function reads
\be
V_{\rm QN}
\simeq
-\frac{2\,\gn\,M}{\pi\,R_{\rm s}}
\left[
1-\frac{\pi\,r^2}{18\,R_{\rm s}^2}
\right]
\ ,
\ee
so that it is bounded and its derivative vanishes for $r=0$ (see Fig.~\ref{VqVn}).
This suggests that gravitational forces remain finite, as we shall show in more details next.
\par
In the classical Schwarzschild spacetime~\eqref{gSch}, the Kretschmann scalar
$R_{\alpha\beta\mu\nu}\, R^{\alpha\beta\mu\nu}\sim R^2\sim r^{-6}$ for $r\to 0$, 
whereas for the above quantum corrected metric we have
\be
R_{\alpha\beta\mu\nu}\, R^{\alpha\beta\mu\nu}
\simeq
R^2
\simeq
\frac{64\,\gn^2\,M^2}{\pi\,R_{\rm s}^2\,r^4}
\ .
\ee
This ensures that tidal forces remain finite all the way to the centre, as can be seen more explicitly from the
relative acceleration of radial geodesics approaching $r=0$, to wit
\be
\frac{\ddot{\delta  r}}{\delta r}
=
-R^1_{\ 010}
\simeq
\frac{8\,\gn^2\,M^2}{9\,\pi^2\,R_{\rm s}^4}
\left(1-\frac{\pi\,R_{\rm s}}{4\,\gn\,M}\right)
\ ,
\ee
where $\delta r$ is the separation between two nearby radial geodesics and a dot denotes again the
derivative with respect to the proper time, like in Eq.~\eqref{geodesic}.
We recall that, in the Schwarzschild spacetime, ${\ddot{\delta  r}}/{\delta r}\sim r^{-4}$, which causes the so-called
``spaghettification'' of matter approaching the central singularity.
The point $r=0$ in the quantum corrected geometry can instead be seen as an integrable singularity~\cite{lukash},
where some geometric invariants still diverge but no harmful effects occur to matter.
\par
One can further compute the effective energy-momentum tensor $T_{\mu\nu}$ from the Einstein
tensor $G_{\mu\nu}$ of the metric~\eqref{Vq} and find the effective energy density 
\be
\rho
=
-\frac{G^0_{\ 0}}{8\,\pi\,\gn}
=
\frac{M}{2\,\pi^2\,r^3}\,
\sin\!\left(\frac{r}{R_{\rm s}}\right)
\ ,
\ee
the effective radial pressure 
\be
p_r
=
\frac{G^1_{\ 1}}{8\,\pi\,\gn}
=
-\rho
\ee
and the effective tension
\be
p_t\
=
\frac{G^2_{\ 2}}{8\,\pi\,\gn}
=
\frac{M}{4\,\pi^2\,r^3}
\left[
\sin\!\left(\frac{r}{R_{\rm s}}\right)
-
\frac{r}{R_{\rm s}}\,\cos\!\left(\frac{r}{R_{\rm s}}\right)
\right]
\ .
\ee
The integrals of these quantities over space are finite and, in particular, one obtains
\be
4\,\pi\,\int_0^\infty
r^2\,\rho(r)\,\d r
=
-4\,\pi\,\int_0^\infty
r^2\,p_r(r)\,\d r
=
M
\ee
and
\be
4\,\pi\,\int_0^\infty
r^2\,p_t(r)\,\d r
=
\frac{M}{2}
\ .
\ee
Another quality of this quantum corrected geometry is that it does not contain a (inner) Cauchy horizon
(whenever there exists the outer event horizon), a property which generalises to electrically
charged black holes~\cite{Casadio:2022ndh}.
This excludes potentially serious casual issues which are often present in regular black hole candidates
(see, {\em e.g.}~Refs.~\cite{regular} and references therein).
\par
In light of the above results, the quantum geometry given by coherent states is consistent 
with a inner matter source that does not collapse to a singularity~\cite{Casadio:2021cbv},
similarly to what is found in the bootstrapped Newtonian approach~\cite{BootN,Casadio:2019cux}.
The latter could therefore provide a compatible effective description of the quantum black hole interior
including (some of the) nonlinearities. 
Moreover, the term inside square brackets in Eq.~\eqref{Vq} induces oscillations around the classical
$V_{\rm N}$ whose effects on test bodies could be observed at $r>\Rh$.
Such oscillations are determined by the (quantum) size $R_{\rm s}$ of the matter source and 
another important observation is that the amplitude of these fluctuations around $V_{\rm N}$ 
in the outer region (for $r>\Rh$) decreases for decreasing values of $R_{\rm s}/\Rh$ (see Fig.~\ref{dVq}).
Therefore, one can always choose (finite) values of $R_{\rm s}/\Rh$ so that the oscillations are too small
to be measured by a distant observer.
On the other hand, one could interpret this effect as a damping of transients as $R_{\rm s}$ shrinks
and the collapse proceeds inside the horizon.
\begin{figure}[t]
\centering
\includegraphics[width=10cm]{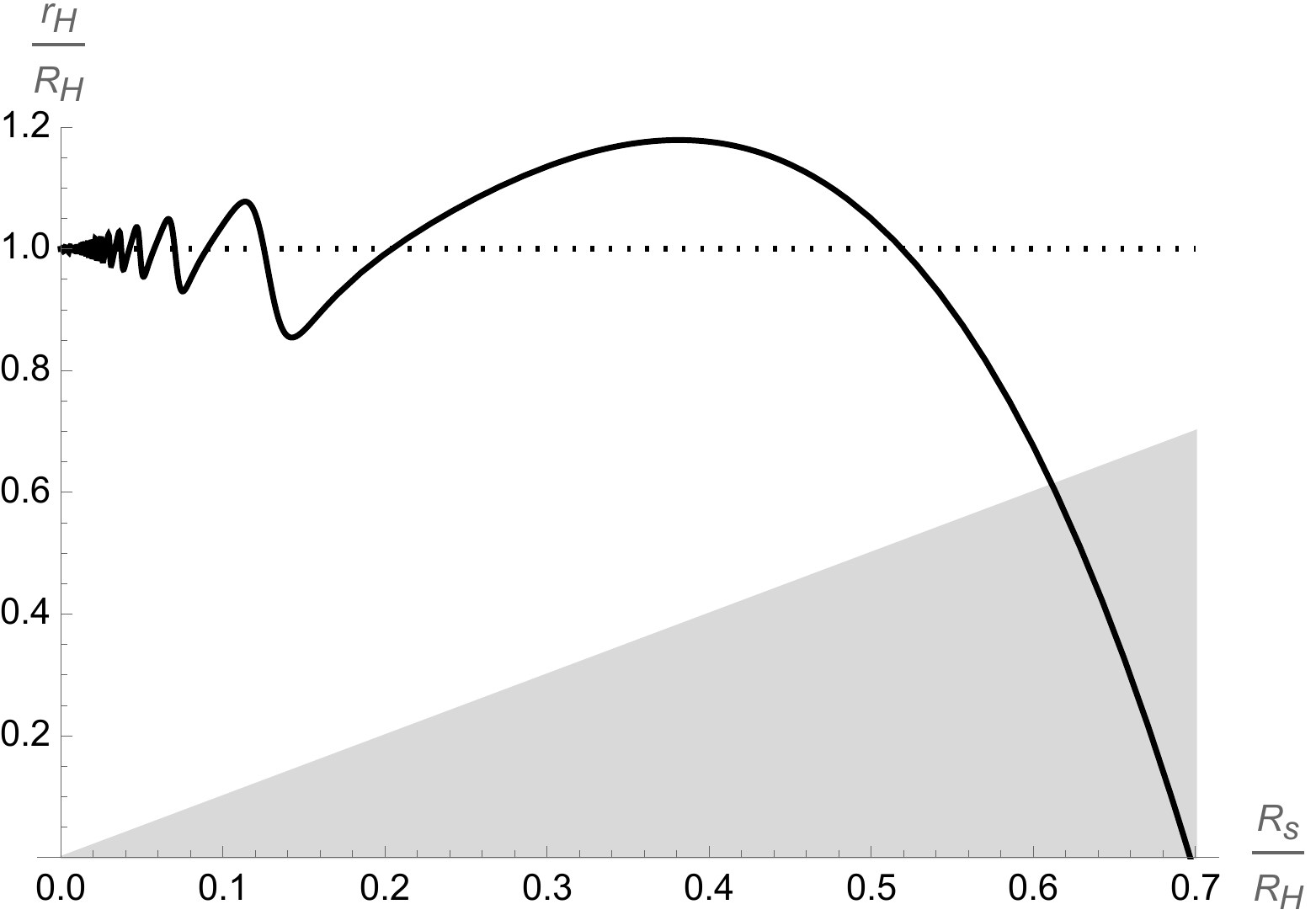}
\caption{Radius $\rh$ of the quantum corrected horizon (solid line) compared to the classical horizon
$\Rh=2\,\gn\,M$ (dotted line) for different values of $R_{\rm s}$.
Shaded regions cover points with $\rh<R_{\rm s}$ and do not correspond to black holes.}
\label{rh}
\end{figure}
\begin{figure}[t]
\centering
\includegraphics[width=10cm]{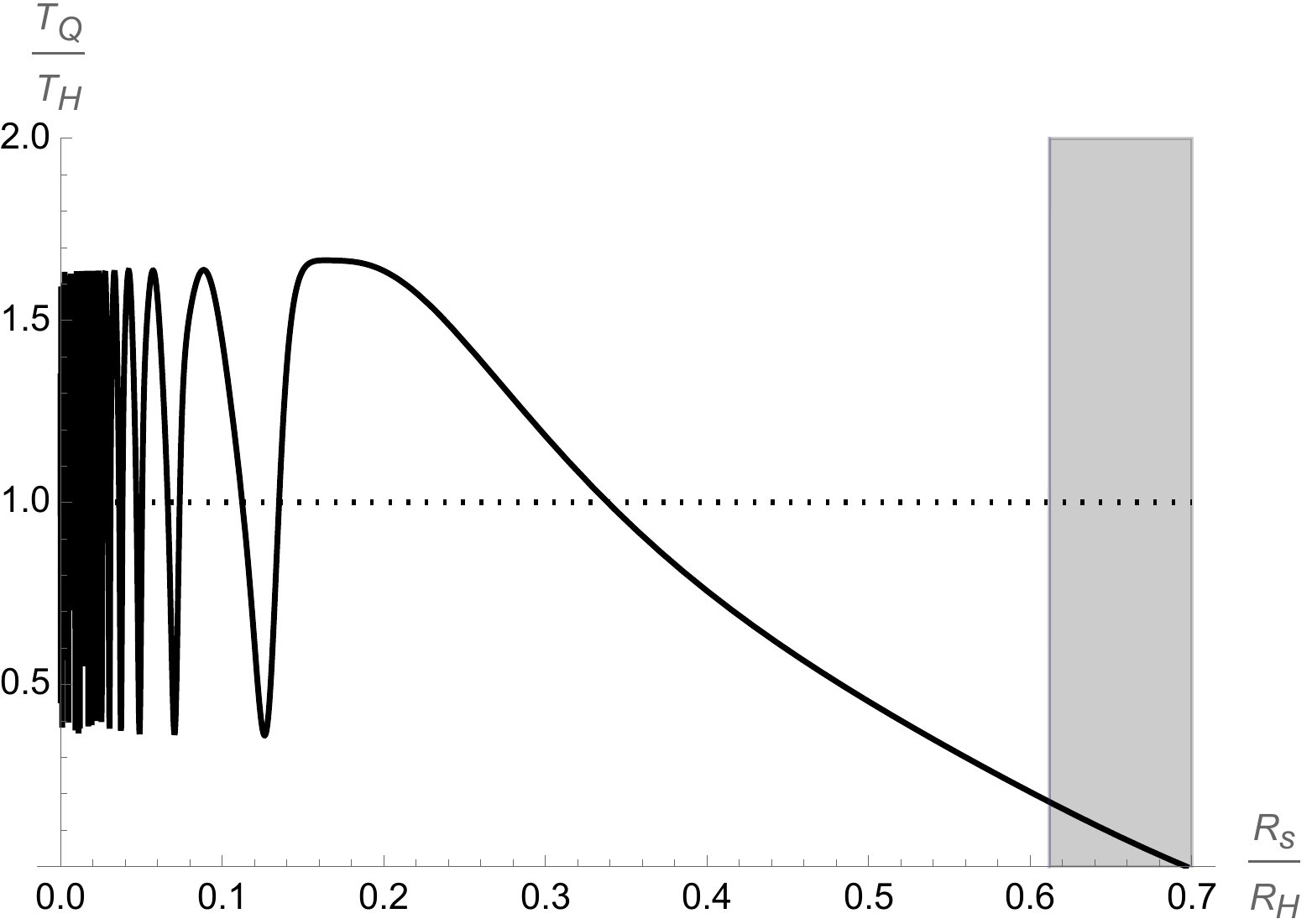}
\caption{Temperature $T_{\rm Q}$ of the quantum corrected black hole (solid line) compared to the
Hawking result $T_{\rm H}=\hbar/4\,\pi\,\Rh$ (dotted line) for different values of $R_{\rm s}$.
Shaded regions cover points with $\rh<R_{\rm s}$ and do not correspond to black holes.}
\label{Th}
\end{figure}
\par
As can be seen from Fig.~\ref{VqVn}, the quantum corrected metric still contains a horizon 
when the cut-off scale $R_{\rm s}\lesssim \Rh$.
The horizon radius $r=\rh$ is given by the solution of $2\,V_{\rm QN}=-1$, which can be computed numerically
for different values of $R_{\rm s}$ (see Fig.~\ref{rh}).
In particular, one can see that $\rh=\Rh$ for very specific values of $R_{\rm s}$, the largest one being
$R_{\rm s}\simeq 0.53\,\Rh$, which is very close to the preferred case obtained in Ref.~\cite{Casadio:2021cbv}.
A remarkable result is that the horizon radius $\rh$ shrinks for $R_{\rm s}$ approaching $\Rh$.
In fact, $\rh=R_{\rm s}$ for $R_{\rm s}\simeq 0.6\,\Rh$ and $\rh$ would further vanish for $R_{\rm s}\simeq 0.7\,\Rh$.
This means that the material core cannot be too close in size to the classical gravitational radius $\Rh$
in order for it to lie inside the actual horizon.
The shaded regions in Fig.~\ref{rh} correspond to values of $\rh < R_{\rm s}$, which are therefore
not black holes.
\subsection{Thermodynamics}
\label{A:thermo}
The above departure of $\rh$ from the Schwarzschild radius $\Rh=2\,\gn\,M$ will also give rise to corrections
for the horizon area $\mathcal A_{\rm H}$ and Bekenstein-Hawking entropy~\cite{bekenstein},
\be
S_{\rm QBH}
=
\frac{\mathcal A_{\rm H}}{4\,\lp^2}
=
\frac{\pi\,\rh^2}{\lp^2}
\label{S_BH}
\ee
and for the black hole temperature~\cite{hawking}
\be
T_{\rm Q}
=
\frac{\hbar\,\kappa}{2\,\pi}
=
\frac{\hbar}{2\,\pi}\left.\frac{\partial V_{\rm Q}}{\partial r}\right|_{r=\rh}
\ ,
\ee
where $\kappa$ is the surface gravity at the horizon.
In particular, Fig.~\ref{rh} shows that the quantum corrected horizon radius $\rh\simeq \Rh$ for
$R_{\rm s}\lesssim \Rh/2=\gn\,M$.
Therefore, the quantum corrected gravitational entropy~\eqref{S_BH} remains very close to its classical
Schwarzschild value $S_{\rm BH}=4\,\pi\,M^2/\lp^2$ as long as the matter core is sufficiently smaller
than $\Rh$.
\par
The quantum corrected Hawking temperature can also be computed only numerically and is
displayed in Fig.~\ref{Th}.
One can then see that quantum corrected black holes are colder then their classical counterpart,
that is $T_{\rm Q}<T_{\rm H}$, if $R_{\rm s}\gtrsim \Rh/2$.
In fact, $T_{\rm Q}$ would vanish for $R_{\rm s}\simeq 0.7\,\Rh$, which precisely correspond
to $\rh=0$ but falls outside the range $\rh>R_{\rm s}$ representing proper black holes.
One could argue that the shaded regions in Fig.~\ref{rh} still represent some intermediate
stage in the gravitational collapse or will perhaps become relevant at the end of the Hawking
evaporation.~\footnote{For the late stage 
of the Hawking evaporation in the corpuscular picture, see {\em e.g.}~\cite{Casadio:2019tfz}
and references therein.} 
\section{Conclusions and outlook} 
\label{S:conc}
\setcounter{equation}{0}
In this work we started from the description of the static and spherically symmetric Schwarzschild geometry
in terms of the coherent state of a massless scalar field on a reference flat spacetime, as previously
employed, {\em e.g.}~in Refs.~\cite{Casadio:2016zpl,Mueck:2013mha,Casadio:2017cdv,Casadio:2020ueb}.
We recall that, in order to grant the coherent state a proper normalisation, we must relax the condition
that its mean field reproduces the classical behaviour everywhere inside the horizon of a black hole~\cite{Casadio:2021onj}.
In fact, it remains questionable whether the mean field approximation should make sense at all in the
interior of a black hole as a (macroscopic) quantum object~\cite{qbh1,qbh2,Casadio:2021cbv,Almeida:2021sci},
but here we took the less drastic viewpoint that a mean field exists, although it cannot reproduce the 
expected classical solution.
\par
The weaker requirement in Eq.~\eqref{QCo} that the coherent state yields the classical
behaviour in the exterior can be achieved by removing modes of wavelength shorter than
a UV cut-off $R_{\rm s}\lesssim \Rh$, where $R_{\rm s}$ can be physically interpreted as the size
of the (quantum) matter source~\cite{Casadio:2021cbv}.
The expectation value of the gravitational potential then displays oscillations around the classical
$V_{\rm N}$, whose amplitudes decrease for shrinking $R_{\rm s}$.
We already commented that the specific shape of such deviations will be affected by
the form of the UV cut-off, but the qualitative dependence on the scale $R_{\rm s}$
should be fairly well captured by the simplest choice of a sharp cut-off in Eq.~\eqref{Vqq}.
One could then interpret this effect as a decay in time due to the relative collapse of the inner
material core of size $R_{\rm s}$ with respect to the outer boundary $R_\infty$ of the Newtonian
region.
Eq.~\eqref{eq:RsRh} then tells us that a core that shrinks down to $R_{\rm s}\simeq\Rh/65$
roughly corresponds to an increase by 26 orders of magnitude in the outer Newtonian
region of size $R_\infty$.
Assuming the last signals generated by the black hole formation (when $R_{\rm s}$ crosses $\Rh$)
are approximately located at the edge of the outer Newtonian potential, and $R_\infty$ expands at
the speed of light, for a solar mass black hole with $\Rh\sim 1\,$km, this means a time of around
$10^{10}$ years, the present age of the Universe.
Such a consideration could be relevant for scenarios which predict a semiclassical bounce~\cite{bounce}.
It is also tempting to push the model towards smaller and smaller black hole masses and conjecture that
the oscillations shown in Fig.~\ref{dVq} grow significantly and totally disrupt the classical picture
for $R_{\rm s}\sim \Rh\sim \lp$.
This would exclude the existence of stable remnants of Planckian size at the end of the Hawking
evaporation~\cite{hawking}, in agreement with other approaches among those in Refs.~\cite{qbh2}.
\par
The simple analysis we presented in this work neglects the precise quantum nature of the matter source
and a better understanding of the interior might be obtained from the bootstrapped Newtonian
picture~\cite{Casadio:2017cdv,BootN,Casadio:2019cux,Casadio:2021gdf} or other modifications of the
gravity-matter coupling~\cite{Feng:2019dwu}.
For a quantum source, the size $R_{\rm s}$ needs not be sharply defined and one should include
the effects of its uncertainty onto the gravitational state, like it was done by considering the bootstrapped
Newtonian potential generated by a source of finite size in Ref.~\cite{Casadio:2020ueb},
or like in Ref.~\cite{Casadio:2021cbv}, where a bound on the compactness was obtained from
the quantisation of the geodesic motion of the surface of a sphere of dust.
The latter result in particular provides a clear motivation for including the UV
scale $R_{\rm s}\sim \gn\,M$ in the definition~\eqref{QCo} of the coherent state of gravitons,
although the use of a sharp cut-off in Eq.~\eqref{Vqq} is justified mostly by the simplicity of the calculation.
Moreover, we note that the natural time evolution of quantum states would imply that the oscillations
around the expected classical potential could make the region around the horizon rather
fuzzy~\cite{Casadio:2013tma,qbh2}, with possibly relevant implications for the causal structure
of astrophysical black holes and their interaction with infalling matter. 
\par
We conclude by remarking that the modified thermodynamics described in Section~\ref{A:thermo}
complements the results from Ref.~\cite{Casadio:2022pla}, in which different forms of information entropy
where computed for the matter core of Ref.~\cite{Casadio:2021cbv}.
In particular, it was found that matter entropy grows with the ADM mass $M$ like $\ln(M^\alpha)$, where 
the power $0.4<\alpha<0.9$.
Since the quantum corrected gravitational entropy~\eqref{S_BH} is still proportional to the horizon area and grows
like $M^2$ for sufficiently small matter cores, the entropy of quantum black holes is expected to be
dominated by the gravitational contribution, like in their classical counterparts.
It will be interesting to investigate whether this conclusion is also supported by a direct calculation of the
information entropy for the coherent states of the geometry~\cite{progress}.
\section*{Acknowledgments}
I would like to thank A.~Giusti and A.~Platania for useful discussions.
I am partially supported by the INFN grant FLAG and my work has also been carried out in
the framework of activities of the National Group of Mathematical Physics (GNFM, INdAM).
\end{document}